\documentclass[a4paper,10pt]{article}
\usepackage[dvips]{graphicx}
\usepackage{amssymb,amsmath}
\numberwithin{equation}{section}

\begin{document}

\title{\bf{Starobinsky model with $f$-essence}}
\author{Shynaray Myrzakul$^{1}$\footnote{Email: shynaray1981@gmail.com},  \,   Kairat Myrzakulov$^{2}$\footnote{Email: krmyrzakulov@gmail.com},  \,  Sabit Bekov$^{2}$\footnote{Email: ss.bekov@gmail.com}, \, Tolkynay Myrzakul$^{1}$\footnote{Email: trmyrzakul@gmail.com} \\ and    Ratbay Myrzakulov$^{2}$\footnote{Email: rmyrzakulov@gmail.com} \\ \textit{$^{1}$Department of Theoretical and Nuclear Physics, Al-Farabi Kazakh National} \\ {University,  Almaty, 050040, Kazakhstan} \\ \textit{$^{2}$Eurasian International Center for Theoretical Physics and  Department of General } \\ \textit{ $\&$  Theoretical Physics, Eurasian National University, Astana 010008, Kazakhstan}}


\date{}
\maketitle
\begin{abstract}
In this paper, we consider a cosmological model of the flat and homogeneous universe for the Starobinsky model  $F(R)=\alpha R + \beta R^{2}$, which is  non-minimally coupled with $f$-essence. For this model we obtained the field equations and considered particular solutions of the coupling and fermionic field functions. It is shown that the fermionic field can describe a nature of the universe.\end{abstract}

\maketitle

\section{Introduction}

As it is known, the main theory describing gravitational phenomena in nature is general theory of relativity (GR). A correctness of this theory is confirmed by various experimental and observational data. However, GR can not fully describe some dynamics of evolution of the universe such as an accelerated expansion of the universe [1,2]. The most acceptable hypothesis for explaining this expansion of the universe is dark energy (DE), but the nature of DE has not been established  yet. At the present time various alternatives of GR have been proposed. One of such alternative theories is the theory of $F(R)$ gravity, where $F$ is some function of Ricci scalar $R$ (see the works [3, 4]). Also,  some cosmological aspects of $F(R)$ gravity with various matter fields were considered in reference [5-9]. The Starobinsky model is one of examples of were the studied $F(R)$ gravity [10]. In referens [11-16] various applications of this theory in cosmology.

In last years, there has been increased interest in cosmological models with fermionic fields [17-21]. Fermionic field in the early era of the evolution of the universe plays the role of an inflaton, in the late era of the evolution of the universe it plays the role of dark energy. Recently,  was proposed model with non-canonical form of kinetic energy for the fermionic field, well-known as $f$-essence (see the works [22-26]). 


In this work, we consider the Starobinsky model non-minimally coupled with $f$-essence for homogeneous and isotropic Friedmann-Robertson-Walker (FRW) metric. The corresponding equations of motion are determined and solution for scale factor in quasi-de Sitter is obtained. Moreover, the cosmological parameters such as Hubble parameter, equation of state parameter and deceleration parameter are found. The results obtained satisfied models of dark energy and able to describe the late time evolution of the universe.

We adopted the signature as $\left(+,-,-,-\right)$ and natural units $8\pi G = c = \hbar = 1$.

\section{Action and equation of motion}

In this section, we consider the Starobinsky model that non-minimally coupled with $f$-essence for the FRW metric. In general, $F(R)$ gravity is expressed through action, as
\begin{equation}
S = \int d^{4} x \sqrt{-g} \left(\frac{1}{2k} F(R) + L_{m}\right),
\end{equation}
where $k= \frac{8\pi G}{c^{4}} $, $g$ is the determinant of the matrix tensor $g_{\mu \nu}$, $F(R)$ is some function of Ricci scalar $R$ and $L_{m}$ is matter Lagrangian. Varying this action by the metric tensor, we obtain the following equation of motion
\begin{equation}
F(R) R_{\mu \nu} - g_{\mu \nu} F(R) + \left[g_{\mu \nu}  - \nabla_{\mu} \nabla_{\nu} \right] F(R) = k T_{\mu \nu},
\end{equation}
here $\mu, \nu$ are indices running from 0 to 3 and $T_{\mu \nu}$ is the energy momentum tensor, which is determined from expression
\begin{equation}
T_{\mu \nu} = - \frac{2}{\sqrt{-g}} \frac{\delta \left(\sqrt{-g} L_{m}\right)}{\delta g^{\mu \nu}}.
\end{equation}

The dependence of the function $F(R)$  on the Ricci scalar in this paper is given similarly to the Starobinsky model $F(R)=\alpha R + \beta R^{2}$, then the action (1) is rewritten as

\begin{eqnarray}
S  = \int d^4  x \sqrt{-g} \left[ h(u) \left(\alpha R + \beta R^{2}\right) + 2 K(Y,u)\right],
\end{eqnarray}
Here for this model $u=\bar{\psi}\psi$ is the biliniar function,  $\psi$ is fermion function and $\bar{\psi}$ is its adjoint function, $h(u)$ is some function, representing the coupling with gravity and fermion field and  $K(Y,u)$  is some modification of Lagrangian of the fermion fields.For $ K (Y, u) = Y - V $ is the standard Lagrangian fermion field.

Together with the action (1), we consider the Friedman-Robertson-Walker metric
\begin{equation}
 \label{FRW}
ds^{2}  = - dt^{2} + a^{2}(t) \left( dx^{2} + dy^{2} + dz^{2} \right),
\end{equation}
where $a(t)$ is a scale factor that depends only on the time $t$. For this metric, we have the following expressions:
$$
\sqrt{-g} = a^{3}, \ R = 6 \left(\frac{\ddot{a}}{a} + \frac{\dot{a}^{2}}{a^{2}}\right), \ Y=\frac{1}{2}i\left(\bar{\psi}\gamma^0\dot{\psi}-\dot{\bar{\psi}}\gamma^0\psi\right),
$$
where a dot on a symbol means differentiation with respect to time $t$. 

Hence for the metric (5), we can be rewritten the action (4) as
\begin{equation}
{\cal S}=\int d^4x\,a^3 \left[h \left(\alpha R + \beta R^{2}\right) - \lambda \left(R - 6 \frac{{\ddot a}}{a} - 6 \frac{{\dot a}^2}{a^2}\right) + 2 K \right]\,.
\end{equation}

By varying this action with respect to $R$, we can determine the Lagrange multiplier $\lambda$ as 
$$
\lambda = \sigma_{i} h \left(\alpha + 2 \beta R \right).
$$
In the case of $\sigma_{1} = 0$, we have $\lambda =0$ or for $\sigma_{2} \neq 0$, then, respectively, $\lambda \neq 0$. The second case corresponds to a cosmological model with a Lagrangian multiplier. In this paper we shall consider the second case.

Consequently, we can write the point-like Lagrangian as
\begin{equation}
L = 6 \alpha a h \dot{a}^{2} + 6 \alpha a^{2} \dot{a}\dot{h}  + \beta a^{3}  h R^{2} + 12 \beta a h R \dot{a}^{2} + 12 \beta a^{2} R  \dot{a}\dot{h} + 12 \beta a^{2} h \dot{a} \dot{R} - 2 a^{3}  K.
\end{equation}
Further we use the Euler-Lagrange and the Hamiltonian constrain equations 
$$
\frac{\partial L}{\partial q_{i}} - \frac{d}{dt} \frac{\partial L}{\partial \dot{q_{i}}} = 0, 
$$
$$
E_{\cal L}= \frac{\partial {\cal L}}{\partial {\dot a}}{\dot a}+\frac{\partial {\cal L}}{\partial {\dot R}} {\dot R}+\frac{\partial {\cal L}}{\partial {\dot \psi}}\dot{\psi}+\frac{\partial {\cal L}}{\partial {\dot{\bar{\psi}}}}\dot{\bar{\psi}}-{\cal L}=0,
$$
where $q_{i}$ are generalized coordinates, depends of variables $a$, $R$, $\psi$ and $\bar{\psi}$. Or in a more detailed form
\begin{equation}
\frac{\partial L}{\partial a} - \frac{d}{dt} \frac{\partial L}{\partial \dot{a}} = 0, 
\end{equation}
\begin{equation}
\frac{\partial L}{\partial R} - \frac{d}{dt} \frac{\partial L}{\partial \dot{R}} = 0, 
\end{equation}
\begin{equation}
\frac{\partial L}{\partial \psi} - \frac{d}{dt} \frac{\partial L}{\partial \dot{\psi}} = 0, 
\end{equation}
\begin{equation}
\frac{\partial L}{\partial \bar{\psi}} - \frac{d}{dt} \frac{\partial L}{\partial \dot{\bar{\psi}}} = 0, 
\end{equation}
\begin{equation}
E_{\cal L}= \frac{\partial {\cal L}}{\partial {\dot a}}{\dot a}+\frac{\partial {\cal L}}{\partial {\dot R}} {\dot R}+\frac{\partial {\cal L}}{\partial {\dot \psi}}\dot{\psi}+\frac{\partial {\cal L}}{\partial {\dot{\bar{\psi}}}}\dot{\bar{\psi}}-{\cal L}=0.
\end{equation}

We can determine the corresponding of the field equations for our model as
\begin{equation}
R = 6 \left(\frac{\ddot{a}}{a} + \frac{\dot{a}^{2}}{a^{2}}\right),
\end{equation}
\begin{equation}
3 \alpha \left(\frac{\dot{a}^{2}}{a^{2}}  + \frac{\dot{a}}{a}\frac{\dot{h}}{h} \right)  + 6 \beta \left( \frac{\dot{a}^{2}}{a^{2}} + \frac{\dot{a}}{a} \frac{\dot{h}}{h}  + \frac{\dot{a}}{a} \frac{\dot{R}}{R} - \frac{1}{12} R \right) R  - \frac{1}{h}\left( Y K_{Y} - K\right) = 0,
\end{equation}
\begin{equation}
\alpha \left( \frac{\dot{a}^{2}}{a^{2}} + 2 \frac{\ddot{a}}{a} + 2  \frac{\dot{a}}{a}\frac{\dot{h}}{h} + \frac{\ddot{h}}{h}\right) + 2 \beta \left[\ddot{R} + 2 \left(\frac{\dot{a}}{a} + \frac{\dot{h}}{h} \right) \dot{R} + \left(\frac{\dot{a}^{2}}{a^{2}} + 2 \frac{\ddot{a}}{a} + 2 \frac{\dot{a}}{a} \frac{\dot{h}}{h} +\frac{\ddot{h}}{h}- \frac{1}{4} R \right) R \right] + \frac{1}{h} K = 0,
\end{equation}
\begin{equation}
K_{Y} \dot{\psi} + 0.5 \left(3 \frac{\dot{a}}{a} K_{Y} + \dot{K_{Y}}\right) \psi - i K_{Y} \gamma^{0}  \psi - 3i \left[\alpha \left(\frac{\dot{a}^{2}}{a^{2}}  + \frac{\ddot{a}}{a}\right) + 2 \beta \left(\frac{\dot{a}^{2}}{a^{2}} + \frac{\ddot{a}}{a}  - \frac{1}{12} R\right)R\right] h_{u} \gamma^{0} \psi =0,
\end{equation}
\begin{equation}
K_{Y} \dot{\bar{\psi}} + 0.5 \left(3 \frac{\dot{a}}{a^{2}} K_{Y} + \dot{K_{Y}} \right) \bar{\psi} + i K_{u} \bar{\psi} \gamma^{0}  + 3i \left[\alpha \left(\frac{\dot{a}^{2}}{a^{2}}  + \frac{\ddot{a}}{a} \right) + 2 \beta \left(\frac{\dot{a}^{2}}{a^{2}} + \frac{\ddot{a}}{a} - \frac{1}{12} R \right)R \right] h_{u} \bar{\psi}  \gamma^{0} = 0. 
\end{equation}

Also, we can rewrite this system of equations in the values of the Hubble parameter $H=\frac{\dot{a}}{a}$ in the following form
\begin{equation}
R = 6 \dot{H} + 12 H^{2},
\end{equation} 
\begin{equation}
3 \alpha \left(H^{2}  + H \frac{\dot{h}}{h} \right)  + 6 \beta \left(H^{2} + H \frac{\dot{h}}{h}  + H \frac{\dot{R}}{R} - \frac{1}{12} R \right) R  - \frac{1}{h}\left( Y K_{Y} - K\right) = 0,
\end{equation}
\begin{equation}
\alpha \left(3 H^{2} + 2 \dot{H} + 2 H \frac{\dot{h}}{h} + \frac{\ddot{h}}{h}\right) + 2 \beta \left[\ddot{R} + 2 \left(H + \frac{\dot{h}}{h} \right) \dot{R} + \left(3H^{2} + 2 \dot{H} + 2 H \frac{\dot{h}}{h} - \frac{1}{4} R \right) R \right] + \frac{1}{h} K = 0,
\end{equation}
\begin{equation}
K_{Y} \dot{\psi} + 0.5 \left(3 H K_{Y} + \dot{K_{Y}}\right) \psi - i K_{Y} \gamma^{0}  \psi - 3i \left[\alpha \left(2 H^{2} + \dot{H}\right) + 2 \beta \left(2 H^{2} + \dot{H}  - \frac{1}{12} R\right)R\right] h_{u} \gamma^{0} \psi =0,
\end{equation}
\begin{equation}
K_{Y} \dot{\bar{\psi}} + 0.5 \left(3 H K_{Y} + \dot{K_{Y}} \right) \bar{\psi} + i K_{u} \bar{\psi} \gamma^{0}  + 3i \left[\alpha \left(2 H^{2} + \dot{H} \right) + 2 \beta \left(2 H^{2} + \dot{H} - \frac{1}{12} R \right)R \right] h_{u} \bar{\psi}  \gamma^{0} = 0. 
\end{equation}

In order to describe the dynamics of our universe, we need from system equations (13)-(17) or (18)-(22) to determine a dependence of the scale factor $a$ and Hubble parameter on the time $t$. However, we can see these systems are higher-order differential equations and from which it following  that not easy to obtained of their analytical solutions. Also, we need to determine the form of the function $h$ and $K$ from equations (13)-(17). In the next section, we will borrow a solution to these problems.

\section{Cosmological solution}

From the metric (5), we shows that the main parameter which describes the dynamics of the evolution of a homogeneous and isotropic universe is a scale factor $a$. The dependence of the scale factor $a$ on the cosmological time can be determined from the system of equations (13) - (17). However, as can be seen, this system consists of a high order nonlinear differential equations, the solution of which is a difficult task, as well as the need to determine the explicit form of the functions $h$ and $K$. In this paper, we confine ourselves to the consideration of particular solutions of these functions. To do this, we define in the beginning the form of the function   $h(u)$ and $K(Y,u)$ in the form
\begin{equation}
h=h_0 u^n, K=K_{o} Y-V_{0} u,
\end{equation}
where $n, \, K_{0}, \, V_{0}$ and $h_0$ are some constants. Further, substituting these solutions into equations (16) and (17), and multiplying both sides of equation (16) by a function $\psi^{\dagger}$ and, respectively, equation (17) is multiplied by a function $\psi$, then equating them and taking into account that we have the following expression 
\begin{equation}
\dot{u} + 3 \frac{\dot{a}}{a} u =0.
\end{equation}
Integrating this equation, we obtain the dependence of the parameter $u$ on the scale factor $a$ as
\begin{equation}
u=\frac{u_0}{a^3},
\end{equation}
where  $u_0$ is a integration constant. 

In the next place
Further, substituting expressions (23) and (25) into equations (13) and (14), we define the following expressions
\begin{equation}
2  a^{2} \dot{a} a^{(3)} + 2 \left(1-3n\right) a \dot{a}^{2} \ddot{a} -a^{2} \ddot{a}^{2}  + \frac{\alpha \left(1 - 3 n\right)}{6 \beta}a^{2}\dot{a}^{2} - 3 \left(2n + 1\right) \dot{a}^{4} - \frac{V_{0} u_{0}^{1-n}}{18 \beta h_{0}} a^{3 n + 1}= 0,
\end{equation}

For $n=1, \  \alpha = 1$ and $\beta = 1$ the equation (26) can be rewrite as
\begin{equation}
2  a^{2} \dot{a} a^{(3)} - 4 a \dot{a}^{2} \ddot{a} - a^{2} \ddot{a}^{2} - \frac{1}{3}a^{2}\dot{a}^{2} - 9 \dot{a}^{4} - C a^{4}= 0,
\end{equation}
where $C= \frac{V_{0} u_{0}^{1-n}}{18\beta  h_{0}}$. To solve equation (29) we confine ourselves to a solution of the form of the de Sitter solution $a=a_0 e^{\xi t}$, then we have the following characteristic equation
\begin{equation}
12 \xi^{4} + \frac{1}{3} \xi^{2} + C = 0,
\end{equation}
solution of this equation
$$
\xi = \pm  \sqrt{\frac{-1 \pm \sqrt{1-432 C}}{72}}.
$$

Therefore, in order for the accelerated expansion of the universe to be fulfilled, it is necessary to specify a condition under which the coefficient in the exponent is positive, that is, consider only a positive value for $ C \leq 0 $, then the determined scale factor will take the form

\begin{equation}
a=a_0 e^{\sqrt{\frac{-1 \pm \sqrt{1-432 C}}{72}} t},
\end{equation}
finally, function $u$ we obtain
\begin{equation}
u = \frac{u_{0}}{a_0^3 e^{ 3 \sqrt{\frac{-1 \pm \sqrt{1-432 C}}{72}} t}}.
\end{equation}

So the Hubble parameter  has 
\begin{equation}
H = \frac{\dot{a}}{a} = \sqrt{\frac{-1 \pm \sqrt{1-432 C}}{72}}=constant.
\end{equation}

Using expression of the parameter of equation of condition $\omega$ and deceleration $q$:
\begin{equation}
\omega = \frac{p}{\rho} , 
\end{equation} 
\begin{equation}
q = - \frac{\ddot{a} a}{{\dot{a}}^2}=\frac{1}{2}\left(1+3\omega\right),
\end{equation} 
we define that
\begin{equation}
\omega = - 1, \ q=-1.
\end{equation}

\section{Conclusions}
\label{cinque}  

Thus, in this paper we have considered some cosmological aspects of the Starobinsky model that interact non-minimally with the f-essence for a flat and homogeneous universe. In the first section, we gave a short introduction in the theory of gravity. In the second section, the Lagrange function (7) was defined for the Friedman-Robertson-Walker metric and using the Euler-Lagrange equations and the hamilton-energy condition determined the corresponding equations of motion (13) - (17). As can be seen, these equations are higher-order nonlinear differential equations, the solution of which is a complicated problem. Also, to solve the system (13) - (17) it is necessary to determine the explicit form of the function $ h (u) $ and $ K (Y, u) $. In the third section we determined the following dependence $ u = \frac {u_ {0}} {a ^ {3}} $, and also considered the following particular solutions $ h = h_0 u ^ n, K = K_ {0} Y - V_ {0} u $. Substituting the values (23), (25) obtained for $ u $, $ h $ and $ K $ to equation (14), we obtained a third-order nonlinear differential equation depending on one variable $ a $ in the form (26). However, the analytic solution of this equation turned out to be a complicated problem and we restricted ourselves to a more compact form of this equation, that is, for the values of the constants $ n = 1, \, \alpha = 1 $ and $ \beta = 1 $ equation (26) takes the compact form (27).
To solve this equation, we considered the de Sitter solution in the form $ a = a_0 e ^ {\ xi t} $, then using the characteristic equation (28) we finally determined the dependence of the scale factor $ a $ on the time $ t $, (29). Further, using this result, we easily determined the Hubble parameter $ H $, the parameter of the equation of state $ \omega $ and the deceleration parameter $ q $, expressions (31) and (34). It is seen from expression (36) that the parameter of the equation of state and the deceleration parameter are minus one, which corresponds to the model of dark energy. This result is useful to describe the accelerated expansion of the modern Universe and not to contradict modern astronomical observational data.


\begin{thebibliography}{}
\bibitem{Perlmutter} Perlmutter S. et al. \textit{Measurements of omega and lambda from 42 high-redshift supernovae}.  The Astophysical Journal, {\bf 517}, N2, 565-586 (1999).   [arXiv:astro-ph/9812133]
\bibitem{Riess} Riess et al. \textit{Observational evidence from supernovae for an accelerating universe and a cosmological constant}. The Astronomical Journal, {\bf 116}, N3, 1009-1038 (1998). [arXiv:astro-ph/9805201]
\bibitem{Peebles} Peebles P.J.E., Ratra B. \textit{The cosmological constant and dark energy}. Reviews of Modern Physics, {\bf 75}, 559-606 (2003). [arXiv:astro-ph/0207347]
\bibitem{Copeland} Copeland E.J., Sami M., and Tsujikawa S. \textit{Dynamics of dark energy}. International Journal of Modern Physics D,  {\bf 15}, 1753 (2006). [arXiv:hep-th/0603057]
\bibitem{Amendola} Amendola L., Tsujikawa S. \textit{Dark energy: Theory and observations}. Cambridge University Press. 491 (2010).
\bibitem{Okabe} Chiba T., Okabe T., Yamaguchi M. \textit{Kinetically driven quintessence}. Physical Review D, {\bf 62}, N2, 3511 (2000). [arXiv:astro-ph/9912463]
\bibitem{Tsujikawa} Tsujikawa S. \textit{Quintessence: A Review}. Classical and Quantum Gravity, {\bf 30}, 214003 (2013). [arXiv:1304.1961]
\bibitem{Khurshudyan} Khurshudyan M.,  Chubaryan E. and  Pourhassan B. \textit{Interacting Quintessence Models of Dark Energy}. International Journal of Theoretical Physics,  {\bf 53}, 2370-2378 (2014). [arXiv:1402.2385]
\bibitem{Caldwell}  Caldwell R.R. \textit{A Phantom Menace? Cosmological consequences of a dark energy component with super-negative equation of state}. Physical Letters B,  {\bf 545}, 23-27 (2002). [arXiv:astro-ph/9908168]
\bibitem{Odintsov} Nojiri S.  and Odintsov S. D. \textit{Inhomogeneous equation of state of the universe: Phantom era, future singularity and crossing the phantom barrier}. Physical Review D, {\bf 72}, 023003 (2005). [arXiv:hep-th/0505215]
\bibitem{Mukhanov1} Armendariz-Picon C., Damour T., Mukhanov V.F. \textit{$k$-inflation}. Physical Letters B,  {\bf 458}, N7, 209-218 (1999). [arXiv:hep-th/9904075]
\bibitem{Mukhanov2} Armendariz-Picon C., Mukhanov V.F., Steinhardt P.J. \textit{Essentials of $k$-essence}. Physical Review D, {\bf 63}, N10, 3510 (2010). [arXiv:astro-ph/0006373]
\bibitem{1011.4337} Myrzakulov R. \textit{Fermionic K-essence}. [arXiv:1011.4337]
\bibitem{Jamil:2011mc} Jamil M., Momeni D., Serikbayev N.S. and Myrzakulov R. \textit{FRW and Bianchi type I cosmology of f-essence}. Astrophysics and Space Science, {\bf 339}, 37 (2012). [arXiv:1112.4472]
\bibitem{1103.5918} Tsyba P., Yerzhanov K., Esmakhanova K., Kulnazarov I., Nugmanova G., Myrzakulov R. \textit{Reconstruction of $f$-essence and fermionic Chaplygin gas models of dark energy}. [arXiv:1103.5918]
\bibitem{1107.1008} Jamil M., Myrzakulov Y., Razina O., Myrzakulov R. \textit{Modified Chaplygin gas and solvable $f$-essence cosmologies}. Astrophysics and Space Science, {\bf 336}. 315-325 (2011). [arXiv:1107.1008]
\bibitem{Kulnazarov:2010an} Kulnazarov I., Yerzhanov K., Razina O., Myrzakul S., Tsyba P. and Myrzakulov R. \textit{G-essence with Yukawa Interactions}. European Physical Journal C, 
{\bf 71}, 1698 (2011). [arXiv:1012.4669]
\bibitem{1203.2804} Bamba K., Razina O., Yerzhanov K., Myrzakulov R.  \textit{Cosmological evolution of equation of state for dark energy in G-essence models}.  International Journal of Modern Physics D. {\bf 22}  1350023 (2013). [arXiv:1203.2804]
\bibitem{1012.5690} Razina O., Kulnazarov I., Yerzhanov K., Tsyba P., Myrzakulov R. \textit{Einstein-Cartan gravity with scalar-fermion interactions}. Central European Journal of Physics, {\bf 10}, N1, 47-50 (2012). [arXiv:1012.5690]
\bibitem{1012.3031} Yerzhanov K.K., Tsyba P.Yu., Myrzakul Sh.R., Kulnazarov I.I., Myrzakulov R.  \textit{Accelerated expansion of the Universe driven by G-essence}. [arXiv:1012.3031]
\bibitem{Buchdahl:1983zz} 
  Buchdahl H.A. \textit{Non-linear Lagrangians and cosmological theory}. Monthly Notices of the Royal Astronomical Society, {\bf 150}, 1-8 (1970).
\bibitem{Nojiri} Nojiri S., Odintsov S.D., Saez-Gomez D. \textit{Cosmological reconstruction of realistic modified $F(R)$ gravities}. Physics Letters B, {\bf 681}, N74, 74-80 (2009). [arXiv:0908.1269]
\end{thebibliography}
\end{document}